\input harvmac
\noblackbox
\def\IZ{\relax\ifmmode\mathchoice
{\hbox{\cmss Z\kern-.4em Z}}{\hbox{\cmss Z\kern-.4em Z}}
{\lower.9pt\hbox{\cmsss Z\kern-.4em Z}}
{\lower1.2pt\hbox{\cmsss Z\kern-.4em Z}}\else{\cmss Z\kern-.4em
Z}\fi}
\def\IB{\relax{\rm I\kern-.18em B}}
\def\IC{{\relax\hbox{$\inbar\kern-.3em{\rm C}$}}}
\def\ID{\relax{\rm I\kern-.18em D}}
\def\IE{\relax{\rm I\kern-.18em E}}
\def\IF{\relax{\rm I\kern-.18em F}}
\def\IG{\relax\hbox{$\inbar\kern-.3em{\rm G}$}}
\def\IGa{\relax\hbox{${\rm I}\kern-.18em\Gamma$}}
\def\IH{\relax{\rm I\kern-.18em H}}
\def\II{\relax{\rm I\kern-.18em I}}
\def\IK{\relax{\rm I\kern-.18em K}}
\def\IP{\relax{\rm I\kern-.18em P}}

\font\cmss=cmss10 \font\cmsss=cmss10 at 7pt
\def\IR{\relax{\rm I\kern-.18em R}}


\lref\polwit{J. Polchinski and E. Witten, ``Evidence for Heterotic-Type I
String Duality,'' Nucl. Phys. {\bf B460} (1996) 525, hep-th/9510169.}
\lref\growth{
J.~McGreevy, L.~Susskind and N.~Toumbas,
``Invasion of the giant gravitons from Anti de Sitter space,''
JHEP {\bf 0006}, 008 (2000), 
hep-th/0003075.
}
\lref\screen{
N.~Kaloper, E.~Silverstein and L.~Susskind,
``Gauge symmetry and localized gravity in M theory,''
hep-th/0006192.}
\lref\edstrong{
E.~Witten,
``Strong Coupling Expansion Of Calabi-Yau Compactification,''
Nucl.\ Phys.\  {\bf B471}, 135 (1996)
hep-th/9602070.
}
\lref\edbaryon{E. Witten, ``Bound States and Branes in Anti-de-Sitter
Space'', 
JHEP {\bf 9807}, 006 (1998)
hep-th/9805112.
}
\lref\uvir{
L.~Susskind and E.~Witten,
``The holographic bound in anti-de Sitter space,''
hep-th/9805114;  
A.~W.~Peet and J.~Polchinski,
``UV/IR relations in AdS dynamics,''
Phys.\ Rev.\  {\bf D59}, 065011 (1999)
hep-th/9809022;
V.~Balasubramanian, P.~Kraus, A.~Lawrence and S.~P.~Trivedi,
``Holographic probes of anti-de Sitter space-times,''
Phys.\ Rev.\  {\bf D59}, 104021 (1999)
hep-th/9808017.
}
\lref\hw{
P.~Horava and E.~Witten,
``Heterotic and type I string dynamics from eleven dimensions,''
Nucl.\ Phys.\  {\bf B460}, 506 (1996)
hep-th/9510209.
}
\lref\memsol{
M.~J.~Duff and K.~S.~Stelle,
``Multi-membrane solutions of D = 11 supergravity,''
Phys.\ Lett.\  {\bf B253}, 113 (1991).
}
\lref\stromvafa{
A.~Strominger and C.~Vafa,
``Microscopic Origin of the Bekenstein-Hawking Entropy,''
Phys.\ Lett.\  {\bf B379}, 99 (1996)
hep-th/9601029.
}
\lref\itzhaki{
N.~Itzhaki, J.~M.~Maldacena, J.~Sonnenschein and S.~Yankielowicz,
``Supergravity and the large N limit of theories with sixteen  supercharges,''
Phys.\ Rev.\  {\bf D58}, 046004 (1998)
hep-th/9802042.
}
\lref\blackp{G. Horowitz and A. Strominger,
``Black strings and P-branes,''
Nucl.\ Phys.\  {\bf B360}, 197 (1991).
}
\lref\savstern{S. Sethi and M. Stern, 
``D-brane bound states redux,''
Commun.\ Math.\ Phys.\  {\bf 194}, 675 (1998)
hep-th/9705046.
}
\lref\matbranes{
T.~Banks, W.~Fischler, S.~H.~Shenker and L.~Susskind,
``M theory as a matrix model: A conjecture,''
Phys.\ Rev.\  {\bf D55}, 5112 (1997)
hep-th/9610043;
T.~Banks, N.~Seiberg and S.~Shenker,
``Branes from matrices,''
Nucl.\ Phys.\  {\bf B490}, 91 (1997)
hep-th/9612157.
}
\lref\flybys{L. Susskind, private communication; S. Giddings,
E. Katz, and L. Randall
``Linearized gravity in brane backgrounds,''
JHEP {\bf 0003}, 023 (2000)
hep-th/0002091.
}
\lref\polsuss{
J.~Polchinski, L.~Susskind and N.~Toumbas,
``Negative energy, superluminosity and holography,''
Phys.\ Rev.\  {\bf D60}, 084006 (1999)
hep-th/9903228.
}
\lref\giddings{S. Giddings, talk at Aspen 2000 String Workshop.}
\lref\barak{ 
B.~Kol,
``Thermal monopoles,''
JHEP {\bf 0007}, 026 (2000)
hep-th/9812021.
}
\lref\banks{ 
T.~Banks,
``Cosmological breaking of supersymmetry or little Lambda goes back to  the future II,''
hep-th/0007146.
}
\lref\ovrut{
A.~Lukas, B.~A.~Ovrut, K.~S.~Stelle and D.~Waldram,
``Heterotic M-theory in five dimensions,''
Nucl.\ Phys.\  {\bf B552}, 246 (1999)
hep-th/9806051;
A.~Lukas, B.~A.~Ovrut, K.~S.~Stelle and D.~Waldram,
``The universe as a domain wall,''
Phys.\ Rev.\  {\bf D59}, 086001 (1999)
hep-th/9803235.
}
\lref\chs{
C.~G.~Callan, J.~A.~Harvey and A.~Strominger,
``Supersymmetric string solitons,''
hep-th/9112030.
}

\Title{\vbox{\baselineskip12pt\hbox{hep-th/0009057}
\hbox{SLAC-PUB-8611}
}}
{\vbox{\centerline{Extended Objects}
\centerline{from} 
\centerline{Warped Compactifications
of M theory}}}
\smallskip

\centerline{
Eva Silverstein
 }
\medskip
\centerline{Department of Physics~~~and~~~ SLAC}
\centerline{~~~Stanford University}
\centerline{~~~Stanford, CA 94305/94309}
\medskip

\noindent

We study the massive spectrum of fully wrapped branes in
warped M-theory compactifications, including
regimes where these states are parametrically lighter than
the Planck scale or string scale.  We show that many such
states behave classically as extended objects in
the noncompact directions  in the sense
that their mass grows with their size
as measured along the Poincare
slices making up the noncompact dimensions.  
On the other hand these states can be quantized in a nontrivial
regime:  in particular their spectrum
of excitations in a limited regime can be obtained by a warped Kaluza-Klein
reduction from ten dimensions.  
We briefly discuss scattering processes and loop effects involving
these states, and also note 
the possibility of an exponential 
growth in the number of bound states of these objects as a function
of energy.

\Date{September 2000}

\newsec{Introduction}

One of the features of the web of M-theory duals is the presence
of various limits in which a distinctive spectrum of states emerges
as the lightest excitation above the (super-) gravity 
modes.  For example in some
limits strings dominate, in others particles including those
coming from wrapped branes.

Compactifications of M theory down to $d$ Poincare-invariant
dimensions are generically warped products,
with $d$-dimensional Minkowski space varying over a compact
$11-d$-manifold $K$.  
That is, the metric is of the form
\eqn\genmet{ds^2=a(y)dx_{||}^2+H_{ij}(y)dy^idy^j}
where $x_{||}^\mu$, $\mu=0,\dots, d-1$ are coordinates
along the Poincare slices and $y_i, i=d, \dots, 10$ are
coordinates along the internal manifold.  
In this paper we will study basic aspects of
the massive spectrum in such compactifications (focusing on
two concrete examples: Horava-Witten theory compactified
on $K3$ \edstrong, and type I' theory as studied
by Polchinski and Witten \polwit\ compactified on a 5-torus).  

We will find objects which behave in the
noncompact $x_{||}$ directions in a way
that is in some sense
intermediate between particles and higher branes.  These
come from ordinary particles or fully wrapped branes in the ten
or eleven-dimensional
description of the physics, but ones for which the mass
and effective size as measured along the Poincare slices
varies over the compactification in such a way that the
energy grows as some positive power of the thickness
$\delta x_{||}$.  We
will refer to such objects as 
elastic states.  By size of
the object we will mean its thickness or the size of the short-distance cutoff
scale, measured with respect to the $x_{||}$ coordinates,
as probed in appropriate scattering experiments.
As we will discuss, in some regimes this is greater than
or equal to the Compton wavelength of our excitations so
that in these cases this is a classically intuitive
accounting of the size scale.      

In a general warped geometry of the form \genmet, the size $R$
of an object as measured in the $x_{||}$ coordinates
($R\equiv \delta x_{||}$) depends
on its characteristic proper size $r_0(y)$ and on the warp factor $a(y)$,
via the relation
\eqn\Rrel{R={1\over\sqrt{a(y)}}r_0(y).}
The energy of an object depends on
$a(y)$ and the metric $H_{ij}(y)$ of
the internal space (as well as the profiles of any additional
fields with nontrivial VEVs in a given background).
As we move the object around $K$ to different values of $y_i$,     
its effective size changes and its energy changes.  For
ordinary point-particle quantum field theory the energy is inversely
related to the size at high energies.\foot{In AdS realizations
of quantum field theory one sees membrane-shaped
objects when a particle falls toward the horizon
in AdS \polsuss\screen\giddings, with their energy decreasing
as their size increases as opposed to the cases we will focus on 
here.}  We will see in our examples
that the size instead grows as some positive power of the energy for
certain massive states in warped compactifications.
This variation in the size of the object results from
a combination of its intrinsic thickness in $11d$ M theory
and the warping of the metric involved in
measuring its size in the $x_{||}$ coordinates.

The phenomenon of growth of the size of states with
energy is generic at high energies due to the
presence of black holes, and is seen in other
contexts in many corners of M theory (see \growth\
for a recent example and references to earlier ones).
One result of our analysis is that warped compactifications
provide another place where this intriguing effect emerges.
Our analysis uses the basic relations arising from
the gravitational redshift in geometries of
the form \genmet\ that also comes into the UV/IR correspondence
developed for AdS space
in studying the AdS/CFT correspondence \uvir; our work 
involves in some ways a generalization of those studies to other
warped geometries where there is no (evident) field theory
dual, but where the energy/size relation can be studied directly
in the gravity theory.  The wrapped branes we study are
in some sense generalizations of the baryon states studied
in backgrounds with quantum field theory duals \edbaryon.

Although the elastic states have a growth in size with energy
reminiscent of branes, their spectrum and interactions 
at long distances can
be calculated via ordinary Kaluza-Klein reduction in
the warped geometry \genmet.  
This spectrum of states, while growing in number faster
with energy than for ordinary unwarped Kaluza-Klein states
on an circle, exhibits
fewer degrees of freedom than a real brane.
In this sense our elastic states are intermediate between
particles and branes (and perhaps analogous to elastic
solids, hence our nomenclature).  
However if we consider the set of potential bound states of 
any number $N$ of elastic states, we
obtain a spectrum with enough degrees of freedom
to describe modes of a continuous medium, and we
discuss at the end a speculation for using these
ideas to construct and quantize real branes using
warped compactifications.  

Direct couplings of objects of different size in this
type of system are suppressed due to their separation
in the internal dimensions, much like in similar systems
studied in the context of the AdS/CFT correspondence.
The size of these objects however does manifest itself in the
cross sections for scattering
processes mediated by electromagnetism and gravity.

One limitation of this approach is that these objects
in the regime we study them here are heavier than
certain Kaluza-Klein modes of massless $11d$ 
supergravity fields.    
So in the $d$-dimensional description, we see one
or more extra dimensions before we see the elastic states.
As we will see, the elastic states can be made very
nearly stable in this same limit, and therefore
at least do not decay into the lighter KK modes.
(In some circumstances it may transpire that there is
an AdS/CFT-like duality between a $d$-dimensional
QFT and the supergravity modes, coupled to the rest
of the system including the elastic states.  Then
the elastic states could be studied consistently in a 
completely $d$-dimensional description of the system.)

Another limitation is in our ability to calculate detailed
physical quantities at the length scale corresponding to
the size of the objects here.  In the Horava-Witten example,
the size is determined by the intrinsice thickness of the
M2-brane, whose proper size is of order $l_{11}$.  We can ameliorate
this problem by considering the classical physics
of a large-N collection of such objects, as we discuss
in \S2\ and \S3.  To discuss the quantum behavior of
systems of this sort, it might be useful to find an
example in which the basic object is a (possibly wrapped)
fundamental string, in a perturbative regime.  



The paper is organized as follows.  In \S2 we will
introduce the Horava-Witten example and
compute the energy/size relations that arises there
(explaining in particular how the size is defined
and determined).  We will work out the scale of masses 
and sizes arising
from Kaluza-Klein reduction.  
In \S3 we work out (in less detail) the same procedure
for type I' theory.
In \S4 we will provide a preliminary discussion of
several interesting aspects and applications of
these results.  In \S4.1 we will comment on the
similarities and differences with branes, and discuss
a possibility for 
an exponentially growing spectrum of
bound states.  In \S4.2\ we discuss scattering
amplitudes and loop effects in these models.  
\S4.3 we conclude
with a discussion of other issues and future directions.    

\newsec{Horava-Witten Example}

\subsec{Mass-Size relation in HW theory}

M theory compactified on $S^1/\IZ_2\times K3$ has 
a warped metric \edstrong
\eqn\hwmet{
ds^2=y^{-1/3} dx_{||}^2
+y^{2/3}(dK^2+{V_0^2\over l_{11}^6}dy^2)
}
where $l_{11}$ is the eleven-dimensional Planck length, 
and $y=c+2\sqrt{2} w$ with
\eqn\wdef{
w={{\pi\rho_0-x^{11}}\over V_0}6\pi\sqrt{2}l_{11}^3(k-12)
}
where $x^{11}$ is a coordinate along the $S^1/\IZ_2$ direction
and $k$ is the number of fivebranes at the 
$x^{11}=0$ end of the interval.
In these formulas, $\pi\rho_0$ is the size of the $S^1/\IZ_2$
as measured with the coordinate $x^{11}$ and $V_0$ is the volume
of the K3 as measured in the K3 metric $dK^2$.  These along
with $c$ are
three of the moduli of the solution (the others being the other
moduli of the K3 metric $dK^2$). 
The result \wdef\ is valid in the regime $V_0 >> l_{11}^4$
and $\rho_0 >> V_0^{1/4}$.  

Consider an M2-brane 
held at a point $y$ in the interval, and
wrapped on a cycle of the K3 with area 
$A(y)=A_0y^{2/3}$.
Because the interval is very large ($\rho_0 >> l_{11}$),
the eleven-dimensional three-form 
gauge potential $C_{MNP}$ has a Kaluza-Klein
excitation with a tunably small mass. This means that our
wrapped M2-brane can be made stable to a good approximation
by taking the interval large enough.      

Its energy, taking into account the warping, is 
\eqn\hwmass{
E=\sqrt{g_{00}}m_0 = y^{-1/6}
\biggl(y^{2/3}{A_0\over l_{11}^3}\biggr)=y^{1/2}{A_0\over l_{11}^3}
}       
where $m_0=y^{2/3}{A_0\over l_{11}^3}$ is the proper energy of the
wrapped M2-brane.

The extent of this wrapped brane along the $x_{||}$ directions
is on the other hand given by \Rrel\
\eqn\hwsize{
R={1\over {\sqrt{a(y)}}}r_0(y)
}
in terms of a characteristic proper size $r_0$.  We will presently
argue that this characteristic size is in fact $l_{11}$.
Taking this value, we obtain from \hwsize\ 
\eqn\hwgenR{
R=l_{11}y^{{1\over 6}}
}

Combining \hwmass\ and \hwgenR, we obtain an energy-size relation
\eqn\hwgenrel{
E=T R^3
}
where the effective tension $T$ is 
\eqn\hwgenten{
T={A_0\over l_{11}^6}
}

It is tempting to infer a mass scale for excitations from
this relation of the order $\tilde\mu\sim T^{1/4}$.  However
the scale of the tension is a priori not the only one in this problem;
the kinetic energy of the elastic state will in general depend
on further parameters.  Therefore we will have to wait for
the KK analysis of \S2.2\ to obtain a determination of the
scale of excitations of this system.

Now let us fix $r_0$ for this example.  
We can do so in two different regimes that
will be of interest.  Firstly, consider the case
where $A(y) >> l_{11}^2$ for all $y$ relevant
to a given process.  In this regime the
supergravity solution for a wrapped membrane has
a profile of characteristic size $l_{11}$ \memsol,
plus small corrections down by powers of $l_{11}^2/A$.
At the scale $l_{11}$, the supergravity/general relativity
approximation breaks down and we expect new structure
at this scale.       
We therefore expect this scale to be seen by appropriate
$11d$ supergravity modes in scattering
processes, and we take it as a natural choice of 
the intrinsic proper thickness $r_0$ of the object.

Let us also consider the case of 
a regime of the K3 moduli space where $dK^2$ describes
a K3 near an ALE singularity, so that there are 2-cycles
in the K3 with area $A_0<<l_{11}^2$, and moreover
consider a range of $y$ for which 
$A(y)<<l_{11}^2$.  (We still take
$V_0>>l_{11}^4$ to be able to continue to use the
results \hwmet\wdef.  Strictly speaking the supergravity
analysis \edstrong\ does not apply if there are sub-Planckian defects
in the manifold, but the warping is caused by excess
fivebranes on one side of the interval scaling up the volume
of the K3, which will presumably still happen in a regime
where the K3 has localized ALE singularities.  The
analogous question has been studied in the Calabi-Yau
threefold case in \ovrut.)    
The UV cutoff of
the gauge theory obtained near an ALE singularity (where
the wrapped M2-branes emerge as Higgsed gauge bosons
of an ADE gauge theory) is $1/l_{11}$.  Therefore $l_{11}$
is the short-distance cutoff of the theory, plus possibly
small corrections going like powers of $A/l_{11}^2$.
This is also the scale of the gauge coupling of the
nonrenormalizable gauge theory obtained at the ALE singularity.
The Coulomb forces between the Higgsed gauge bosons will 
go like $1/r^4$ for separations $r$ much greater than the
cutoff $l_{11}$.  The form factor for scattering of
these states will then have a characteristic size of order $l_{11}$.  
So in this case also, we will
take $l_{11}$ as the proper size 
$r_0$ of our M2-brane states.  

Finally, for a collection of $N$ M2-branes, we will take
$r_0$ to be $l_{11}N^{1/6}$, which is the characteristic
scale appearing in the supergravity
solution.  
In \S2.3\ we will study these regimes in more detail
given the behavior of the KK wavefunctions we determine in
the next subsection.

\subsec{HW Kaluza-Klein Analysis}

We will now compute the spectrum of Kaluza-Klein
modes of the wrapped M2-branes in the geometry \hwmet.
Let us consider for simplicity here the case of an M2-brane
wrapped on a genus zero cycle of K3.  For these the normal
bundle is the line bundle ${\cal O}(-2)$ (which has no
sections), so the M2-brane cannot move on the K3 without
cost in energy.  Here we will focus on the motion of
the wrapped branes in the $y$ direction, which dominates
over the motion in the K3 in appropriate regimes of moduli.
Let us for simplicity also consider the scalar components
of the wrapped M2-branes.

In order to calculate the spectrum of excitations of these wrapped
M2-branes, we can use a Kaluza-Klein reduction of the action
\eqn\hwkkac{
S_{HW}=\int d^6x_{||}dy \sqrt{G}\biggl[
-G^{MN}\partial_M\phi\partial_N\phi -m_0^2(y)\phi^2
\biggr]
}
Here the indices $M,N$ run over the seven dimensions
of the warped interval, and $G_{MN}$ is given by the corresponding
components of \hwmet, so that $G_{\mu\nu}=\eta_{\mu\nu}y^{-1/3}$
and $G_{yy}=y^{2/3}{V_0^2\over l_{11}^6}$.  There no 
explicit dependence
in the action on the $y$-dependent K3 volume $V(y)=V_0 y^{4/3}$ since
the curve is isolated; there is dependence on the cycle area
$A(y)=A_0y^{2/3}$ through the mass term in \hwkkac.    

The modes are given by solutions of the equation of motion
\eqn\hweom{
\partial_M\biggl(\sqrt{G}G^{MN}\partial_N\phi\biggr)
=\sqrt{G}m_0^2(y)\phi
}   

This becomes, upon plugging in our background, 
\eqn\hwnexteom{y^{-1/3}\partial^2_{x_{||}}\phi
+{l_{11}^6\over V_0^2}\partial_y(y^{-4/3}\partial_y\phi)
=y^{2/3}{A_0^2\over l_{11}^6}\phi
}
Letting $\phi=e^{ik_{||}x_{||}}\tilde\phi$ we have
that $\partial^2_{x_{||}}\phi=-k_{||}^2\phi=\mu^2\phi$ where
$\mu$ is the mass of the excitation in six dimensions.

Defining $\eta=y^{-2/3}\phi$, this becomes
\eqn\hsimpl{
\eta''-\biggl[{{10}\over {9 y^2}}
+ ay^2-by\biggr]\eta =0
}
where
\eqn\abdef{
a={{A_0^2 V_0^2}\over l_{11}^{12}} ~~~~~~~ 
b={{\mu^2 V_0^2}\over l_{11}^6}
}

This is related to a non-relativistic
quantum mechanics problem with effective potential
\eqn\eompot{
V(z)=ay^2-by+{{10}\over {9y^2}}.
}
We are interested in the discrete set of values of $\mu^2$
(which comes into the parameter $b$ \abdef) for which this
quantum mechanics problem has a state with zero energy eigenvalue.

For simplicity let us work in a regime where the last
term in \eompot\ can be dropped.  The problem then
reduces to a harmonic oscillator potential.
We will be interested in determining the mass scale $\mu$ and
the locations 
$y_c$ where the solutions $\eta$ are peaked.
We will then check for self-consistency of this approximation.

In this regime, \hsimpl\ can be rewritten
\eqn\complsq{
-\eta''+a\biggl(y-{b\over{2a}}\biggr)^2\eta
={b^2\over{4a}}\eta
}
The energy eigenvalues in the corresponding harmonic oscillator
problem are given by 
\eqn\enharm{
E_n={b_n^2\over{4a}}=\sqrt{a}(n+{1\over 2})
}
Using this and \abdef\ we find a tower of masses
\eqn\hwmuscale{
\mu_n^2={{2A_0^{3/2}}\over{V_0^{1/2}l_{11}^3}}\sqrt{(n+{1\over 2})}
={1\over \lambda_C^2}
}
where we have indicated the Compton wavelength determined
by this mass scale in the last step.  

The nontrivial locations of the peaks of the wavefunction
are at the length scale
\eqn\peakz{
y_n={b\over a}={{2l_{11}^3\sqrt{n+{1\over 2}}}
\over {A_0^{1/2}V_0^{1/2}}}
}
The values \peakz\ are consistent with our assumption that
the $1/y^2$ term in \eompot\ could be ignored relative
to the linear and quadratic terms for large enough $\mu^2$.

From \peakz\ and \hwgenR, we can determine the effective size of these
excitations:
\eqn\hwRn{
R_n\sim y_n^{1/6}l_{11}
=\mu_n^{1/3}{l_{11}^2\over A_0^{1/3}}
}
They satisfy a mass-size relation
\eqn\kkensize{
\mu_n \sim T R_n^3}
with a tension $T$ agreeing with that found in our analysis in
\S2.1.  

Since these excitations have
a nontrivial size $R_n$, the point-particle Kaluza-Klein 
analysis we have done breaks down at this scale, which
is to say that the mode solutions of \hweom\ apply only
for momentum $k_{||}$ along the Poincare slices which is
smaller than the scale $1/R_n$.  

Note that these states grow (in number) faster than
for example Kaluza-Klein modes on a circle.  Of course the states
of a single elastic state grow much more slowly than those of
say a perturbative string.  In \S4.1\ we will discuss briefly
possible bound states of these objects, whose density of states does
appear to grow exponentially with some power of the energy.

\subsec{Size Scales and Regimes of Moduli}

In order for \wdef\ to be reliable, we
need $V_0 >> l_{11}^4$.  We can then consider different regimes
of area $A(y)$ and M2-brane number $N$.  
Our arguments at the beginning of this section
fixing $r_0=l_{11}$ involved assuming $A(y)$ to be either very large
or very small relative to $l_{11}^2$ for the range
of $y$'s of interest.  Let us check here
that that can be arranged for our KK excitations \hwmuscale\peakz.
At the typical $y$-values $y_n$, we have
\eqn\typarea{
A(y_n)={{2A_0^{2/3}l_{11}^2 (n+{1\over 2})^{1/3}}
\over{V_0^{1/3}}}
}

Suppose we wish to arrange for $A(y)>>l_{11}^2$.  From
\typarea\ this requires 
\eqn\classcase{
A_0^{2/3}(n+{1\over 2})^{1/3}>> V_0^{1/3}
}
In this regime, the membrane is wrapped on a very large
2-cycle, and is therefore heavy and  behaves classically.
Indeed, the condition \classcase\ is the same as the
condition that the Compton wavelength be much
smaller than the object:  $\lambda_C << R_n$.  
When \classcase\ is satisfied for all $n$, the K3 is rather skew
in shape, since in particular 
$\biggl({A_0\over l_{11}^2}\biggr)
>>\biggl({V_0\over l_{11}^4}\biggr)^{1/2}$.
In this regime the elastic states behave classically.

On the other hand, suppose we wish to consider the case
where $A(y)<<l_{11}^2$.  Then we need
\eqn\qmcase{
A_0^{2/3}(n+{1\over 2})^{1/3} << V_0^{1/3} 
}  
For this regime, the Compton wavelength is large compared to
the size scale $R_n$, and the objects behave very quantum
mechanically.  When the K3 is near an ALE singularity, we
argued at the beginning of this section that the cutoff
scale and gauge coupling scale $l_{11}$ determines the
size of the object in this regime as well.   

In both regimes we have studied here, for a small number
$N$ of branes the size $R$ we have been discussing is determined
by the warping from a fundamental proper size $r_0=l_{11}$.
Although we expect structure at this scale as discussed above,
this being the scale of the profile of the supergravity
solutions for M-branes for example, we do not have direct control
over the details of processes at this 
scale.\foot{Perhaps a more tractable generalization would
be a case involving a (possibly wrapped) perturbative string.
In such a case one could calculate reliably scattering
processes at the relevant size scale ($l_s$).}
To study the
elastic states' physics in a classical regime where we
do have control, we can introduce
a large number $N$ of wrapped M2-branes.  Then as discussed
above, $r_0\sim N^{1/6}l_{11}$.  As in studies of
black hole systems (\stromvafa\ etc.) the branes then become
amenable to a classical analysis in certain regions of
the solution \itzhaki.  

Of course in adding branes to 
increase $N$, we increase both the mass and the size of the
object as a function of $N$.  Our point here is that in
addition to that well-known 
effect, at fixed $N$ there is 
the effect we have identified here:
the stretching of
the object along $x_{||}$ due to the warping as
one increases its energy.
Considering a fixed large $N$ allows
us to study the classical physics of the elastic state effect
in a regime where the physics at the scale of the object
is under control.    

We have thus obtained a consistent description of the wrapped
M2-branes in the HW geometry which shows that they grow in
size as a function of their energy.  In this sense they behave
like extended objects.  
We will begin a study in \S4\ 
of the question of the extent to which these objects and their bound
states might behave as familiar extended objects such as branes.

\newsec{The Type I' Case}

In this section we will do a similar analysis of NS5-branes
in the type I' geometry derived in \polwit.  We will
work this case out a little more schematically than the last one,
since the procedure is hopefully clear, but it is worth
exhibiting a second example of the basic effect under consideration.

By some rescalings of the moduli and coordinates 
defined in \polwit\ we can
write the string-frame metric and dilaton schematically in the form
\eqn\Ipmet{ds^2_{string}={\gamma^{5\over 3}\over{(\beta+w)^{1\over 3}}}
(dx_{||}^2+l_s^2dw^2)
}
\eqn\dil{e^\phi={1\over{[\gamma(\beta+w)]^{5\over 6}}}
}      
where $\beta$ and $\gamma$ are some combinations of the
dilaton and radial modulus of the type I' theory.
Here $x_{||}$ denotes the coordinates along the $9d$ 
Poincare slices in this geometry.  

Let us compactify 5 of the spatial $x_{||}$ dimensions on
a $T^5$.  For simplicity let us take the simple square shape
\eqn\torid{
x_{||}^{5,\dots, 9}\cong x_{||}^{5,\dots, 9}+R_0
} 

The thickness of $N$ NS5-branes in the remaining $x_{||}$ directions
can be determined in a large $N$ limit from the supergravity
solution.  This gives \chs
\eqn\Ipsize{
r_0\sim l_s N^{1/2}
}

We can determine the energy-size relation from the warped metric
and dilaton as follows.  Let us work in terms of a
coordinate $y\equiv w+\beta$.  Taking into account 
the variation of the volume of the $T^5$ and the variation
of the dilaton as a function of $y$, the energy is
\eqn\Ien{
E=\sqrt{g_{00}}m_0
={R_0^5\over l_s^6}\gamma^{20/3}y^{2/3}
}
The size is
\eqn\Isize{
R={y^{1/6}\over\gamma^{5/6}}l_sN^{1/2}
}

Putting these together leads to a growth in size with energy:
\eqn\Ireln{
E = T R^{4}
}
with effective tension
\eqn\Iten{
T={{R_0^5\gamma^{10}}\over{l_s^{10}N^2}}
}

The Kaluza-Klein analysis in this case can be done similarly
to that for the Horava-Witten case.

\newsec{Disussion:  Bound States, Scattering, and Other Issues}

\subsec{Remarks on Bound States and Branes}

Though our elastic states have a growth in size with energy
somewhat reminiscent of branes, they each individually only
have a Kaluza-Klein theory's worth of states.  However 
bound states of elastic states might exist (establishing
or ruling out this possibility would require a careful
analysis such as was done for D0-branes \savstern).  If so, these 
could have
a much faster (exponential) growth in number with energy.

In the HW case, the wrapped M2-branes experience gravitational
attraction and gauge ($C_{MNP}$) repulsion.  Because the
$S^1/\IZ_2$ direction is compact, the relevant modes of
$C_{MNP}$ are slightly massive, as discussed in \S2.  
Therefore at very long distances the different elastic states
attract each other.  This suggests bound states might be
possible, though an analysis of the short-distance
structure is required to establish (or rule out) this possibility.

Consider a bound collection of bound states of $N_n$ elastic states at
level $n$ in the spectrum \hwmuscale\ derived in \S2.   
An upper bound on the energy of such a state 
(in that it does not take into account the binding energy) is
\eqn\hwbounden{
\mu_{tot}\sim \sum_n N_n \mu_n
=\sum_n N_n (n+{1\over 2})^{1/4} 
{{\sqrt{2} A_0^{3/4}}\over{V_0^{1/4}l_{11}^{3/2}}}
}
This spectrum grows exponentially.

If bound states do exist here, it is tempting to speculate
that a large number of elastic states could mock up an effective
brane (perhaps in a somewhat analogous way to the way D0-branes 
mock up higher branes in matrix theory \matbranes).
It is further tempting to speculate that this approach can
then provide a new way to quantize effective branes and
study their interactions in some nontrivial regime.

\subsec{Remarks on Scattering Amplitudes}

Given the growth in size with energy of the objects we
have been studying, it is interesting to consider where
this effect would arise in scattering amplitudes.
Contact interactions between large and small elastic states
(which are particles at different points $y_1, y_2$ in the
interval) are suppressed by the separation between
$y_1$ and $y_2$.  As in the case of objects
going by each other in the AdS bulk \flybys, in the
$d$-dimensional description our objects
appear to pass right through each other.\foot{One possibility
is that the large objects form a ring instead of a filled-in
ball in the $x_{||}$ dimensions as in some states in
AdS \polsuss\giddings.}      

However if we consider the electromagnetic and gravitational
interactions of our objects, their size is evident at tree level.
In particular, the thickness of the object leads to a form
factor with characteristic scale $R$ in gauge and gravity-mediated
scattering processes.  

It will be interesting to study the effects of elastic states
and bound states of elastic states in loops.  For example it
will be interesting to calculate the contribution these
objects and their bound states make to the vacuum energy.
For this we need
to know the spectrum of bound states as well as a controlled
prescription for describing their interactions.  It may be
very useful to consider an analogous case where the elastic states
originate as strings (with a warping which amplifies their
$l_s$-scale size).  

\subsec{Discussion}

There are many issues to explore further with these states
(and their many cousins in other warped compactifications).
Generalizations which could be important for a potential
brane picture include cases where the wrapped branes
oscillate in more than one direction of a warped
compactification.  As discussed above, a case which
would give more tractable calculations would be one in which
the basic object is a perturbative string whose mass
and size get warped by the compactification so as to give
an energy-size relation of the kind we have studied here.

One question involves the relevance of AdS/CFT ideas
and results to more generic warped compactifications.
In some ways the elastic states we have discussed here 
are a generalization of the baryon states
identified in AdS/CFT duals and their generalizations \edbaryon.
The growth in size of states with energy is characteristic
of gravity in various high-energy regimes, but can also
occur for some collective excitations in ordinary quantum field theory
(and exponential growth of the density of states can
also occur there in certain regimes \barak). 
It would be interesting to classify the behavior of states
at high energy in QFT vs gravity in this regard, and
in particular to understand if the class of states studied
here is characteristic of gravity or could occur
in a theory with a completely quantum field-theoretic dual.    

Another interesting application of warped geometries
is potentially to cosmology (obtained in appropriate
cases by viewing the direction along which the warping appears
as time $t$).  In such a situation, the wrapped branes can become 
zero-action instantons at 
a singularity occuring at $t=0$ in the cosmology.
It would be interesting to understand whether 
the corresponding instanton sum accounts for (and 
is related to a resolution of) the
initial singularity in this sort of setup.

Particularly in the context of gravity, the effects of
nonlocality are potentially important for many problems
(see e.g. \banks\ for a recent application).
It would be interesting to understand
what role if any these extended states play in 
the nonlocality of gravity in warped compactifications.

\medskip
{\centerline {\bf Acknowledgements}}

I would like to thank A. Adams, 
O. Aharony, N. Arkani-Hamed, T. Banks, M. Berkooz, 
S. Kachru, A. Lawrence, J. Maldacena, M. Peskin, S. Shenker, 
and A. Strominger for useful discussions.
This work is supported by a
DOE OJI grant, by the A.P. Sloan
Foundation, and
by the DOE under contract DE-AC03-76SF00515.

\listrefs

\end